\documentclass[dvips]{article}
\usepackage{icrctc07}

\newcommand{\Gcenter}[2]{
  \dimen0=\ht\strutbox%
  \advance\dimen0\dp\strutbox%
  \multiply\dimen0 by#1%
  \divide\dimen0 by2%
  \advance\dimen0 by-.5\normalbaselineskip
  \raisebox{-\dimen0}[0pt][0pt]{#2}}%

\title{Future plan for observation of cosmic gamma rays in the 100 TeV energy region \\
with the Tibet air shower array : physics goal and overview
}
\shorttitle{Future plan of the Tibet experiment}
\authors{
The Tibet AS$\gamma$ Collaboration\\
M.~Amenomori$^{1}$, X.~J.~Bi$^{2}$, D.~Chen$^{3}$, S.~W.~Cui$^{4}$,
Danzengluobu$^{5}$, L.~K.~Ding$^{2}$, X.~H.~Ding$^{5}$, C.~Fan$^{6}$,
C.~F.~Feng$^{6}$, Zhaoyang Feng$^{2}$, Z.~Y.~Feng$^{7}$,
X.~Y.~Gao$^{8}$, Q.~X.~Geng$^{8}$, H.~W.~Guo$^{5}$, H.~H.~He$^{2}$,
M.~He$^{6}$, K.~Hibino$^{9}$, N.~Hotta$^{10}$, Haibing~Hu$^{5}$,
H.~B.~Hu$^{2}$, J.~Huang$^{11}$, Q.~Huang$^{7}$, H.~Y.~Jia$^{7}$,
F.~Kajino$^{12}$, K.~Kasahara$^{13}$, Y.~Katayose$^{3}$,
C.~Kato$^{14}$, K.~Kawata$^{11}$, Labaciren$^{5}$, G.~M.~Le$^{15}$,
A.~F.~Li$^{6}$, J.~Y.~Li$^{6}$, Y.-Q.~Lou$^{16}$, H.~Lu$^{2}$,
S.~L.~Lu$^{2}$, X.~R.~Meng$^{5}$, K.~Mizutani$^{13,17}$, J.~Mu$^{8}$,
K.~Munakata$^{14}$, A.~Nagai$^{18}$, H.~Nanjo$^{1}$,
M.~Nishizawa$^{19}$, M.~Ohnishi$^{11}$, I.~Ohta$^{20}$,
H. Onuma$^{17}$, T.~Ouchi$^{9}$, S.~Ozawa$^{11}$, J.~R.~Ren$^{2}$,
T.~Saito$^{21}$, T.~Y.~Saito$^{22}$, M.~Sakata$^{12}$,
T.~K.~Sako$^{11}$, M.~Shibata$^{3}$, A.~Shiomi$^{9,11}$,
T.~Shirai$^{9}$, H.~Sugimoto$^{23}$, M.~Takita$^{11}$,
Y.~H.~Tan$^{2}$, N.~Tateyama$^{9}$, S.~Torii$^{13}$,
H.~Tsuchiya$^{24}$, S.~Udo$^{11}$, B.~Wang$^{8}$, H.~Wang$^{2}$,
X.~Wang$^{11}$, Y.~Wang$^{2}$, Y.~G.~Wang$^{6}$, H.~R.~Wu$^{2}$,
L.~Xue$^{6}$, Y.~Yamamoto$^{12}$, C.~T.~Yan$^{11}$, X.~C.~Yang$^{8}$,
S.~Yasue$^{25}$, Z.~H.~Ye$^{15}$, G.~C.~Yu$^{7}$, A.~F.~Yuan$^{5}$,
T.~Yuda$^{9}$, H.~M.~Zhang$^{2}$, J.~L.~Zhang$^{2}$,
N.~J.~Zhang$^{6}$, X.~Y.~Zhang$^{6}$, Y.~Zhang$^{2}$, Yi~Zhang$^{2}$,
Zhaxisangzhu$^{5}$ and X.~X.~Zhou$^{7}$
}

\shortauthors{M. Amenomori et al.}
\afiliations{\noindent
$^{1}$Department of Physics, Hirosaki University, Hirosaki 036-8561, Japan.
$^{2}$Key Laboratory of Particle Astrophysics, Institute of High Energy Physics, Chinese Academy of Sciences, Beijing 100049, China.
$^{3}$Faculty of Engineering, Yokohama National University, Yokohama 240-8501, Japan.
$^{4}$Department of Physics, Hebei Normal University, Shijiazhuang 050016, China.
$^{5}$Department of Mathematics and Physics, Tibet University, Lhasa 850000, China.
$^{6}$Department of Physics, Shandong University, Jinan 250100, China.
$^{7}$Institute of Modern Physics, SouthWest Jiaotong University, Chengdu 610031, China.
$^{8}$Department of Physics, Yunnan University, Kunming 650091, China.
$^{9}$Faculty of Engineering, Kanagawa University, Yokohama 221-8686, Japan.
$^{10}$Faculty of Education, Utsunomiya University, Utsunomiya 321-8505, Japan.
$^{11}$Institute for Cosmic Ray Research, University of Tokyo, Kashiwa 277-8582, Japan.
$^{12}$Department of Physics, Konan University, Kobe 658-8501, Japan.
$^{13}$Research Institute for Science and Engineering, Waseda University, Tokyo 169-8555, Japan.
$^{14}$Department of Physics, Shinshu University, Matsumoto 390-8621, Japan.
$^{15}$Center of Space Science and Application Research, Chinese Academy of Sciences, Beijing 100080, China.
$^{16}$Physics Department and Tsinghua Center for Astrophysics, Tsinghua University, Beijing 100084, China.
$^{17}$Department of Physics, Saitama University, Saitama 338-8570, Japan.
$^{18}$Advanced Media Network Center, Utsunomiya University, Utsunomiya 321-8585, Japan.
$^{19}$National Institute of Informatics, Tokyo 101-8430, Japan.
$^{20}$Tochigi Study Center, University of the Air, Utsunomiya 321-0943, Japan.
$^{21}$Tokyo Metropolitan College of Industrial Technology, Tokyo 116-8523, Japan.
$^{22}$Max-Planck-Institut f\"ur Physik, M\"unchen D-80805, Deutschland.
$^{23}$Shonan Institute of Technology, Fujisawa 251-8511, Japan.
$^{24}$RIKEN, Wako 351-0198, Japan.
$^{25}$School of General Education, Shinshu University, Matsumoto 390-8621, Japan.
}

\email{tsako@icrr.u-tokyo.ac.jp}

\abstract{
The Tibet air shower array, which has an effective area of 37,000 square meters and is located at 4300 m 
in altitude, has been observing air showers induced by cosmic rays with energies above a few TeV.
We are planning to add a large muon detector array to it for the purpose of increasing its sensitivity 
to cosmic gamma rays in the 100 TeV (10 - 1000 TeV) energy region by discriminating 
them from cosmic-ray hadrons.
We report on the possibility of detection of gamma rays in the 100 TeV energy region 
in our field of view, based on the improved sensitivity of our air shower array deduced from the 
full Monte Carlo simulation.
}

\begin{document}
\maketitle
\section{Introduction}
Although SNRs are theoretically considered to be the most plausible candidates 
for acceleration of cosmic-ray hadrons up to PeV energies, no observations 
have succeeded in identifying them so far.
Since accelerated electrons have difficulty producing very high-energy 
gamma rays with energies above 100 TeV via bremsstrahlung or 
inverse Compton scattering, it can be an effective way of obtaining 
clear evidence for hadronic acceleration to detect high-energy 
gamma rays above 100 TeV generating via the decay of neutral pions produced in interactions 
of accelerated hadrons with ambient material, e.g. molecular clouds. 
In other words, study of the energy-dependent morphology of the sources in the 100 TeV region (10 - 1000 TeV)
is a clue to disentangle their acceleration mechanism.

The HESS group reported in 2005 on the discovery of 14 new TeV gamma-ray sources in the southern hemisphere
and many of them have a harder energy spectrum (indices: -1.8 to -2.8) \cite{HESS}, 
which implies that hadrons should be accelerated there. 
No definite claims have been made yet, however, because it is difficult for HESS 
to measure the sources above $\sim$10 TeV.
Also, many of the 14 sources are faint in other wavelengths. 
This strongly suggests that an unbiased survey irrelevant to observations in other wavelengths
using a wide field-of-view apparatus, e.g. an air shower array, is needed.

\section{The Tibet Air Shower Experiment}
The Tibet air shower experiment has been successfully operating at Yangbajing
(90$^{\circ}$31' E, 30$^{\circ}$06' N; 4,300 m a.s.l.) in Tibet, China since 1990.
Being expanded several times, the Tibet air shower (AS) array now consists of 789 plastic scintillation
counters placed on a 7.5 m square grid with an effective area of 37,000 m$^2$, and is detecting
high-energy ($>$ a few TeV) cosmic gamma rays and hadrons.

Using the Tibet AS array, we so far have observed a new cosmic-ray anisotropy in the Cygnus region 
at multi-TeV energies \cite{science} as well as TeV gamma rays from the Crab \cite{crab}, 
Mrk501 \cite{501} and Mrk421 \cite{421}.
Because the anisotropy that we observed in the Cygnus region is relatively narrow, it may favor
the existence of a gamma-ray component as claimed by the Milagro group \cite{mgro2}.
We can not draw any clear conclusion on the region, however, because the AS array at present is unable to 
discriminate gamma rays from hadrons in the multi-TeV energy region. 

\section{The Tibet Muon Detector Array}
\begin{figure}[t]
  \begin{center}
    \includegraphics [bb=0 80 590 620, width=0.48\textwidth]{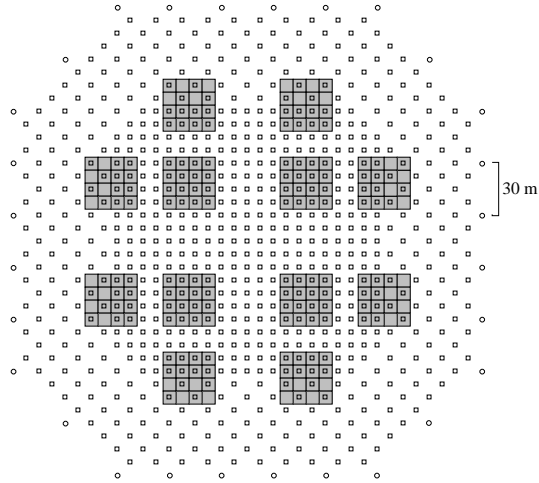}
  \end{center}
  \caption{Schematic view of the Tibet AS+MD array. Open squares and open circles represent 
the surface scintillation detectors that compose the Tibet AS array.
Note that the AS array drawn here is upgraded from the current version so that its effective area 
becomes 50,000 m$^2$ by modifying the configuration of the scintillation detectors. 
Filled squares show the proposed Tibet MD array 2.5 m underground.}\label{config}
\end{figure}
The currently proposed configuration of the Tibet MD array is shown in Figure \ref{config}.
It is composed of 12 pools, each of which consists of 16 cells.
Each cell is a waterproof concrete tank which is 7.2 m wide $\times$ 7.2 m long $\times$ 1.5 m deep 
in size. 
Two 20 inch-in-diameter photomultiplier tubes (PMTs, Hamamatsu R3600) are put on its ceiling, 
facing downwards. 
Its inside is painted with white epoxy resin to waterproof and to efficiently reflect 
water Cherenkov light, which is then collected with the PMTs.
The MD array is set up 2.5 m underground ($\sim$19 radiation lengths)
in order to detect the penetrating muon component of air showers, suppressing the electromagnetic one.
Its total effective area amounts to 9,950 m$^2$ for muon detection with the energy threshold of 
approximately 1 GeV.

\section{Results and Discussions}
We performed detailed Monte Carlo simulations regarding air showers and the responses of soil, 
the AS and MD array, which are described in \cite{myposter}.
Based on the number of collected photoelectrons by the PMTs of the MD array, background hadrons can be
rejected by selecting muon poor events.
Around 100 TeV, the number of hadron-induced events are suppressed down to 0.01\% or less, while 
gamma-induced events are retained by more than 83\%.

Table \ref{perform} summarizes the comparison of performance between HESS and the Tibet AS+MD array.
One can see the advantages of an air shower array, i.e. 
a wide field-of-view ($\sim$1.5 sr) and high duty cycle ($\sim$90\%). 
Although the angular and energy resolution of the AS+MD array around 100 TeV are 
two times worse than that of HESS, the difference is not so critical.
Taking RX J1713 as an example, a famous source in the southern sky with radius of 0.6$^{\circ}$ and 
spectral index of $\sim-$3, the AS+MD array could extract meaningful physical quantities.

\begin{figure}
  \begin{center}
    \includegraphics [bb=50 70 555 530,width=0.48\textwidth]{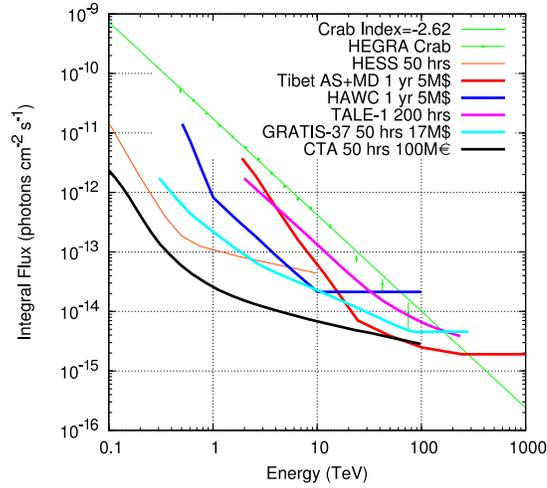}
  \end{center}
  \caption{The attainable integral flux sensitivity (5$\sigma$ in one calendar year) of the 
Tibet AS+MD array to a point-like gamma-ray source, 
together with the sensitivity of HESS and some other future experimental plans.
}
\label{sens}
\end{figure}

The sensitivity of the Tibet AS+MD array to a gamma-ray point source deduced from the simulation
is shown in Figure \ref{sens}.
Its 5$\sigma$ sensitivity in one calendar year will reach 7\% and $\sim$20\% Crab 
above 20 and 100 TeV respectively, and surpass the existing IACTs above 20 TeV.
Furthermore, it may surpass the sensitivity of the next generation IACTs above 40 TeV.
Since no extensive search has been conducted with an apparatus with sensitivity comparable to HESS 
in the northern sky, the AS+MD array could discover as many ($\sim$10) unknown TeV gamma-ray sources 
as HESS did in \cite{HESS}.
\begin{table}
  \caption{Comparison of performance between HESS and the Tibet AS+MD array}
  \label{perform}
  \begin{center}
    \begin{tabular}{|l|c|c|}
      \hline
   &   Tibet AS+MD & HESS \\
   &   $>\sim$100 TeV & $\sim$TeV \\
      \hline
Location   &  30N - 90E & 23S - 16E \\
FOV   &  $\sim$1.5 sr & $\sim$0.02 sr \\
Duty cycle   &  $\sim$90\% & $\sim$10\% \\
Angular res.   &  $\sim0.2^{\circ}$ & $\sim0.1^{\circ}$ \\
Energy res.   &  $\sim$40\% & $\sim$20\% \\
BG rejection   &  $\sim10^{-4}$ & $\sim10^{-2}$ \\
\hline
S/N     &  50/0.3 ev. & 2500/2000 ev.\\
(RX J1713, in  &  $>$100 TeV & $>$1 TeV \\
\cline{2-3}
0.6$^{\circ}$ radius,   &  120/3 & 20/20 \\
1yr or 50hrs) &  $>$40 TeV & $>$40 TeV \\
\hline
      \end{tabular}
  \end{center}
\end{table}

\begin{figure}[h]
  \begin{center}
    \includegraphics [bb=40 70 555 550, width=0.48\textwidth]{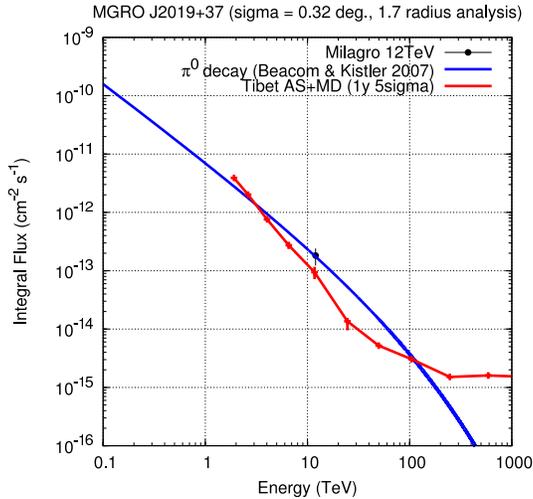}
  \end{center}
  \caption{ 5$\sigma$ sensitivity of the Tibet AS+MD array in one calendar year (red) to a diffuse 
gamma-ray source \cite{mgro}. The blue line is expected from the Cygnus region in the galactic plane. 
}
\label{MGRO}
\end{figure}

As shown in Figure \ref{MGRO}, the Tibet AS+MD array will have more advantage in observing spatially 
diffuse gamma-ray sources, e.g. MGRO J2019+37, discovered by the Milagro experiment \cite{mgro}.

In brief summary, 
among the known sources in our field-of-view, 
Crab, TeV J2032+4130, HESS J1837-069, Mrk 501, Mrk 421, MGRO J2019+37, MGRO J2031+41 and MGRO J1908+06
\cite{mgro2} are sufficiently detectable.
HESS J1834-087, LS I+61 303, Cas A, and M87 are marginal.

\newpage
\section{Acknowledgements}
The collaborative experiment of the Tibet Air Shower Arrays has been
performed under the auspices of the Ministry of Science and Technology
of China and the Ministry of Foreign Affairs of Japan. This work was
supported in part by Grants-in-Aid for Scientific Research on Priority
Areas (712) (MEXT), by the Japan Society for the Promotion of Science,
by the National Natural Science Foundation of China, and by the
Chinese Academy of Sciences.


\bibliographystyle{plain}

\end{document}